\documentclass[amsmath,superscriptaddress,showpacs,prd]{revtex4}

\usepackage{graphicx}

\begin{document}
\title{Can the Mechanism for $\pi_1\to \eta\pi,\eta'\pi$ Hybrid Decays  be Detected?}
\author{Ailin Zhang}
\email{zhangal@theory1.usask.ca}
\affiliation{Department of Physics and Engineering Physics, University of
Saskatchewan, Saskatoon, SK, S7N 5E2, Canada}
\affiliation{Institute of Theoretical Physics, P.O.\ Box 2735, Beijing,
100080,
P.R.\ China}
\author {T.G.\ Steele}
\email{Tom.Steele@usask.ca} 
\affiliation{Department of Physics and Engineering Physics, University of
Saskatchewan, Saskatoon, SK, S7N 5E2, Canada}

\date{\today}

\begin{abstract}
Two mechanisms for the  $\pi_1$  ($J^{PC}=1^{-+}$)  hybrid meson decay processes
$\pi_1\to\eta\pi,\eta'\pi$ are investigated.  These mechanisms are applied to
$\phi\to\eta\gamma,\eta'\gamma$ and $J/\psi\to\eta\gamma,\eta'\gamma$ decays to
illustrate the validity of the decay mechanisms and to obtain independent information on the 
coupling of $\eta,\eta'$ to quark and gluonic operators.   
From this information, we find that
$\Gamma(\pi_1\to\eta\pi)/\Gamma(\pi_1\to\eta'\pi)$ is substantially different in the two
decay mechanisms, and hence future experimental measurements of this ratio will
provide valuable  information for determining 
the mechanism for these hybrid decays.
\end{abstract}


\maketitle
Hybrid meson states, which contain a gluonic degree of freedom in addition to a 
$q\bar q$ pair, can have exotic $J^{PC}$ quantum numbers. Experimental 
evidence for the existence of  two such exotic isovector 
 $\pi_1$ ($1^{-+}$) states is summarized in \cite{PDG}.
The $\pi_1(1400)$, formerly known as the $\hat \rho(1405)$ (mass 
$1376\pm 17\,{\rm MeV}$   and width $300\pm40\,{\rm MeV}$  \cite{PDG}), was observed by both E852 and 
Crystal Barrel in very different production processes decaying into 
$\eta\pi$ \cite{e852,crys}. The $\pi_1(1600)$ (mass $1596^{+25}_{-14}\,{\rm MeV}$, width
$312^{+64}_{-24}\,{\rm MeV}$ \cite{PDG})
was observed by E852 in  
 $\rho\pi$ and $\eta'\pi$ decay channels \cite{e852}. 
The branching ratios for all these channels are not yet measured, but as will be shown below, they can provide important information
for subsequently detecting their dominant decay mechanism. 

Theoretically, the features of hybrids have been studied with the MIT bag 
model \cite{bag}, flux tube models \cite{flux}, potential models \cite{potential}, quark-gluon 
constituent model \cite{constituent,const2}, 
QCD sum rules \cite{sum,width}, lattice simulations \cite{lattice}, and other 
methods \cite{other}, but 
the exploration is neither complete nor definitive. In particular, predictions 
for the hybrid decay widths exhibit some disagreement with the experimental 
results.

Hybrids can
possess exotic $J^{PC}$ quantum numbers such as $0^{+-}$, $1^{-+}$ and $2^{+-}$ which are distinct from those of conventional
$q\bar q$ mesons. As such,  these exotic hybrids have no mixing with
other conventional hadrons which provides an advantage in the
investigation and detection of these states. 
The predicted mass of the exotic $1^{-+}$ hybrid is approximately
 $2.0\,{\rm GeV}$  in lattice simulations \cite{lattice}, while in QCD 
sum rules   the resulting mass prediction is  $1.4\mbox{--}2.1 
\,{\rm GeV}$ \cite{sum, width}.  The sum-rule predictions are slightly
lower than those of the lattice, but are consistent with 
experiment. However, considering the possible accuracy of 
 the sum-rule and lattice calculations, any apparent deviation between the predicted and observed $\pi_1$ masses is insufficient
to assist in the interpretation of the observed states.  Thus it is important to study 
decay features  because they are more 
sensitive to the nature of the $\pi_1$ states.      

The decay modes and relevant decay widths of the  exotic $1^{-+}$ hybrid have been
 studied using QCD sum-rules for three-point correlation functions, but the scheme employed   
for $\eta$-$\eta'$ mixing has a significant effect on the  $\pi_1\to\pi\eta,\pi\eta'$ widths.
In the traditional singlet-octet mixing scheme, these widths are found to be small
\cite{width}, and are similar to those found from selection rules \cite{const2,select}. 
These predictions seem inconsistent with the experimental observations. 
However, in a different  $\eta$-$\eta'$ mixing scheme  \cite{mixing,pennington},
the three-point sum-rule analysis results in an enhancement 
in the $\pi_1\to\eta\pi$ width \cite{tom}.

Enhancement of the $\pi_1\to\eta\pi$ width in the  three-point sum-rule 
analysis clearly indicates the importance of the composition of the $\eta,\eta'$ system.  
However,
there are a number of issues that complicate the sum-rule analyses of the 
$\pi_1$ decays. For example, there is phenomenological evidence that the  
$\eta$-$\eta^\prime$ system has a gluonic component \cite{ball,shifman}, which would clearly 
have an effect on the sum-rule analyses.  Furthermore, the necessary three-point functions have 
only been calculated at the symmetric Euclidean point which leads to a single Borel 
transformation instead of the double transformation needed for a full analysis.
Finally, the traditional three-point sum-rule method obscures  determination of
the dominant hadron decay mechanism, indicating the need for further investigation. 

Hybrids are a many body system containing a quark, anti-quark and gluon ($\bar qqg$), 
which complicates the determination of the allowed  $J^{PC}$ values. To get some feel
for what is involved, consider these complexities in the MIT bag
model. In this model, the gluon in hybrids may be in two different
modes ($TM(1^{--})$ or $TE(1^{+-})$), and  the quark anti-quark pairs may 
also take on different $J^{PC}$ configurations. As a consequence, there 
exist many kinds of quantum number combinations. For example, when the 
quark and antiquark have no relative orbital angle momentum in the pair, 
the $J^{PC}$ of this pair may be $0^{-+}$ or $1^{--}$. Therefore, besides 
the normal $q\bar q$  meson $J^{PC}$  quantum numbers, hybrids may also have exotic $J^{PC}$ such as 
$0^{+-}$, $1^{-+}$ and $2^{+-}$. These kinds of exotic states simplify 
both the theoretical investigation and experimental detection. Among these 
exotic states, the hybrid $\pi_1$ ($1^{-+}$) is predicted as the lowest-lying state in the 
hybrid spectrum. Its $J^{PC}$ construction is: $1^{--}\times 1^{+-}$, {\it i.e.,} 
the gluon is in transverse electric mode $TE(1^{+-})$. 

There are a number of models to describe the
 strong decay of hadrons. Many models describe the strong decay of a meson to a  two-meson final state through creation of a 
quark-antiquark pair which then  combines with the 
quark and antiquark of the original meson to form the two-meson final state. 
However, at a more detailed level,  the quark 
pair creation process can be viewed as either a 
$^3S_1(1^{--})$ or $^3P_0(0^{++})$ intermediate state.

\begin{figure}[hbt]
\centering
\includegraphics[scale=0.5]{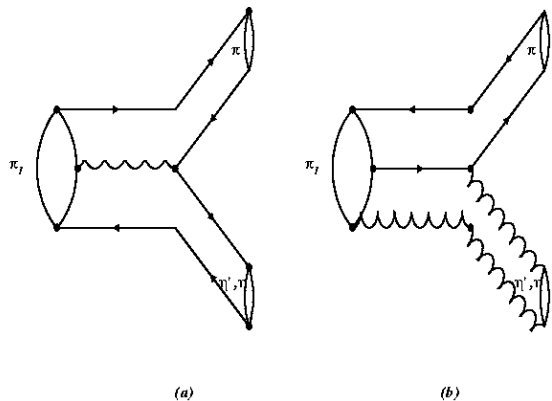}
\caption{
The decay $\pi_1 \to\pi\eta',\eta$
 via the $\bar q q$ mechanism is shown in (a),
while decay via the $gg$ mechanism is shown in (b).  
}
\label{decay_fig}
\end{figure}

The special role of the constituent 
gluon in hybrids leads to some different decay possibilities. In the constituent parton picture, two main decay 
mechanisms, denoted by $\bar q q$ and $gg$ will be studied in this paper. In 
the $\bar q q$ process the gluon 
in the initial hybrid becomes a quark anti-quark pair, while in the $gg$ process  a new gluon is emitted from 
an original constituent quark and combine with the original constituent gluon 
into a final meson. For the $\pi_1\to\pi\eta',\eta$ decays,   these two 
decay mechanisms are illustrated  in Figure \ref{decay_fig}.

It is difficult  to accurately describe  or calculate these decays directly from QCD because of the fundamentally non-perturbative 
nature of strong decays. In this paper
we show that  it is possible to extract the  $\eta,\eta'$ couplings to 
quark and gluonic currents from existing experimental data on  the decays 
$J/\psi\to \gamma\eta',\gamma\eta$ and $\phi\to \gamma\eta',\gamma\eta$, and to apply this information 
to two mechanisms for $\pi_1\to \pi\eta',\pi\eta$ decays.  
This approach avoids the issues that have complicated the sum-rule analyses, and allows determination 
of the dominant hybrid decay mechanism.

\section{Gluonic couplings of $\eta$, $\eta'$ through $J/\psi\to\eta\gamma,\eta'\gamma$ decays }
To extract the $\eta,\eta'$ couplings to gluonic currents, consider the  processes $J/\psi\to\eta\gamma,\eta'\gamma$.
In these processes the photon is emitted by one of the $c$ quarks before their annihilation into lighter quark pairs \cite{ball,shifman}.
The resulting ratio of decay rates occurring if the 
$J/\psi\to\eta\gamma,\eta'\gamma$
decay processes occur  through $\bar cc\to gg\to 
\eta,\eta'$ is
\begin{equation}
{\Gamma(J/\psi\to\eta\gamma)\over \Gamma(J/\psi\to\eta'\gamma)}\simeq
\left|{\langle 0|G\tilde{G}|\eta\rangle \over \langle
0|G\tilde{G}|\eta'\rangle}\right|^2\left({1-m^2_{\eta}/m^2_{J/\psi}\over
1-m^2_{\eta'}/m^2_{J/\psi}}\right)^3,
\end{equation}
where it has been assumed that the $gg$ pair is sufficiently hard so that 
the use of the local operator $G\tilde G=1/2\epsilon^{\mu\nu\lambda\rho}G^a_{\lambda\rho}G^a_{\mu\nu}$ extracted from the $gg$ pair 
is a good approximation.
The experimental value of this decay-width ratio is \cite{bes} 
\begin{equation}
{\Gamma(J/\psi\to\eta\gamma)\over \Gamma(J/\psi\to\eta'\gamma)}
= 0.200\pm0.023.
\label{psi_data}
\end{equation} 
Using the known masses
$m_{J/\psi}=3.1$ GeV, $m_{\eta}=0.548$ GeV, and $m_{\eta'}=0.958$ GeV, the following relation
\begin{eqnarray}
{\langle 0|G\tilde{G}|\eta\rangle \over \langle 
0|G\tilde{G}|\eta'\rangle}\simeq 0.404\pm0.023
\label{GtildeG}
\end{eqnarray}
is obtained.  The uncertainty given in (\ref{GtildeG}) is only associated with the experimental value (\ref{psi_data}). 

The agreement between the value  (\ref{GtildeG})  and the corresponding sum-rule estimates
\cite{pennington,shifman} provides support for the assumed 
 $\bar cc\to gg\to 
\eta,\eta'$ 
mechanism and associated approximations used for these decays.  
However,  the sum-rule analyses are based on specific  models of the 
$\eta$-$\eta'$ system, while the extraction  (\ref{GtildeG}) is independent of 
such considerations and would remain valid with the addition of a  
gluonic component mixing with $\eta$-$\eta'$.  For this reason we will use  (\ref{GtildeG}) in our subsequent analyses.

\section{Quark couplings of $\eta$, $\eta'$ through $\phi\to\eta\gamma,\eta'\gamma$ decays}
The $\phi\to\eta\gamma,\eta'\gamma$ decays must occur through a different mechanism than the $J/\psi$ decays, since  
 a  $\bar ss\to gg\to \eta,\eta'$  process similar to the $J/\psi$ decays considered earlier would lead to  
\begin{equation}
{\Gamma(\phi\to\eta'\gamma)\over \Gamma(\phi\to\eta\gamma)}\simeq
\left|{\langle 0|G\tilde{G}|\eta'\rangle \over \langle
0|G\tilde{G}|\eta\rangle}\right|^2\left({1-m^2_{\eta'}/m^2_{\phi}\over
1-m^2_{\eta}/m^2_{\phi}}\right)^3\simeq 2.8\times 10^{-2},
\end{equation}
where the mass $m_{\phi}=1.02$ GeV and (\ref{GtildeG}) has been used. This value is an order of magnitude larger than the
known experimental value \cite{kloe}
\begin{equation}
{\Gamma(\phi\to\eta'\gamma)\over \Gamma(\phi\to\eta\gamma)}= 4.7\pm 0.47\pm 0.31\times 10^{-3},
\label{phi_data}
\end{equation}
indicating that the  $\bar ss\to gg\to \eta,\eta'$ process does not properly describe the $\phi\to\eta\gamma,\eta'\gamma$ processes. Since  a high energy scale
is needed to approximate the $gg$ mechanism as the coupling of a gluonic operator to mesons, 
it is not surprising that the $gg$ mechanism cannot be applied to $\phi$ radiative decays to 
$\eta,\eta'$ \cite{rad}.

Alternatively,  consider the  $\bar ss\to \bar 
qq\to \eta,\eta'$ mechanism, which could include the direct process 
 $\bar ss\to \eta,\eta'$.
Although the detailed decay mechanism is unknown, this process can be modeled through  $\eta,\eta'$ couplings to quark currents.
Considering the nature of the  the initial $\phi$ state and final $\eta,\eta'$ states, the appropriate 
quark current is of the form $\bar q i\gamma_5 q$ indicating a general  flavour structure.  Note that a coupling to an axial vector 
current would contain a (dominant) anomaly term which would then lead to the gluonic current couplings as ruled out in the above analysis.
The resulting ratio of decay rates in this general  $\bar ss\to \bar 
qq\to \eta,\eta'$ mechanism is 
\begin{equation}
{\Gamma(\phi\to\eta'\gamma)\over \Gamma(\phi\to\eta\gamma)}\simeq
\left|{\langle 0|\bar q i\gamma_5 q|\eta'\rangle \over \langle
0|\bar qi\gamma_5 q|\eta\rangle}\right|^2\left({1-m^2_{\eta'}/m^2_{\phi}\over
1-m^2_{\eta}/m^2_{\phi}}\right)^3.
\label{phi_qq}
\end{equation}
  Using (\ref{phi_data}), we obtain
\begin{equation}\label{ga1}
\left\vert
{\langle 0|\bar qi\gamma_5 q|\eta\rangle \over \langle 0|\bar qi\gamma_5 
q|\eta'\rangle}\right\vert\simeq 0.984\pm0.082 ,
\end{equation}
where the uncertainty only reflects the experimental value (\ref{phi_data}).  

To disentangle the actual flavour structure occurring in the estimate (\ref{ga1}), we turn to the theoretical estimates 
\begin{gather}\label{ga2}
\left\vert{\langle 0|\bar si\gamma_5s|\eta\rangle \over
\langle 0|\bar si\gamma_5s|\eta'\rangle}\right\vert\simeq 0.76\pm 0.10
\\
\left\vert{\langle 0|\bar ni\gamma_5n|\eta\rangle \over
\langle 0|\bar ni\gamma_5n|\eta'\rangle}\right\vert\simeq 2.5
\label{nn_res}
\end{gather}
as  obtained in a recent QCD sum-rule analysis \cite{pennington} (see also \cite{mixing} for other estimates of this quantity), where
$n$ denotes the non-strange $u,d$ quarks in the $SU(2)$ limit. 
Comparison of   (\ref{ga1}), (\ref{ga2}) and (\ref{nn_res}) indicates that the direct $\bar s s\to \eta\eta'$  process is predominant  in  
$\phi\to\eta\gamma,\eta'\gamma$, and the agreement between  (\ref{ga1}) and (\ref{ga2}) validates the approximations used to obtain
(\ref{phi_qq}).  In particular, this agreement indicates that the coupling of the final states to  operators ignored in obtaining (\ref{phi_qq}) must be small enough such that the pseudoscalar current dominates the process.

It is important to note that the numerical value (\ref{nn_res}) is almost unchanged between $\eta,\eta'$ mixing schemes, while 
(\ref{ga2})  shows some scheme dependence.
  Thus we can consider our result (\ref{ga1}) as a mixing-scheme independent extraction of the ratio of the couplings to strange 
pseudoscalar currents, and can safely use the theoretical value for non-strange currents, obviating the absence of 
experimental data that could be used to extract the non-strange ratio.

As a final demonstration of the consistency of our analysis, we return to the $J/\psi$ decays under the assumption of a 
 $\bar cc\to \bar 
qq\to \eta,\eta'$ mechanism, resulting in 
\begin{equation}
{\Gamma(J/\psi\to\eta\gamma)\over \Gamma(J/\psi\to\eta'\gamma)}\simeq
\left|{\langle 0|\bar q i\gamma_5 q|\eta'\rangle \over \langle
0|\bar qi\gamma_5 q|\eta\rangle}\right|^2
\left({1-m^2_{\eta}/m^2_{J/\psi}\over
1-m^2_{\eta'}/m^2_{J/\psi}}\right)^3\gtrsim 1.19,
\end{equation}
where (\ref{ga1}) has been used to obtain a lower bound. This is clearly an inadequate description of the decay process because of its disagreement with the
experimental value (\ref{psi_data}), illustrating that different mechanisms and the resulting coupling of $\eta$, $\eta'$ to different
operators are occurring in each case.  Indeed, if a common decay mechanism existed in the cases considered so far, then the matrix elements
of the relevant operator  would
cancel in the following double ratio resulting in 
\begin{equation}
\frac{\frac{\Gamma(J/\psi\to\eta\gamma)}{\Gamma(J/\psi\to\eta'\gamma)}}{\frac{\Gamma(\phi\to\eta\gamma)}{\Gamma(\phi\to\eta'\gamma)}}
\simeq\frac{\left(1-m_\eta^2/m_{J/\psi}^2\right)^3}{\left(1-m_{\eta'}^2/m_{J/\psi}^2\right)^3}
\frac{\left(1-m_{\eta'}^2/m_{\phi}^2\right)^3}{\left(1-m_{\eta}^2/m_{\phi}^2\right)^3}
=5.59\times 10^{-3}
\end{equation} 
where  masses have been inserted to obtain the numerical value.  By comparison, the experimental value of the double ratio ratio 
obtained from (\ref{psi_data}) and (\ref{phi_data}) is
\begin{equation}
\frac{\frac{\Gamma(J/\psi\to\eta\gamma)}{\Gamma(J/\psi\to\eta'\gamma)}}{\frac{\Gamma(\phi\to\eta\gamma)}{\Gamma(\phi\to\eta'\gamma)}}
=0.2\cdot 4.7\times 10^{-3}=9.4\times 10^{-4} ,
\end{equation}
demonstrating that  the scales associated with the $\phi$ and $J/\psi$ decays must be described by couplings to 
different operators, with the
$J/\psi$ decays best described by coupling to gluonic operators through the  $\bar cc\to \bar 
gg\to \eta,\eta'$ mechanism, while the $\phi$ decays are best described by a coupling to quark operators through the  
 $\bar ss\to \eta,\eta'$   mechanism.

\section{Hybrid  $\pi_1\to\eta\pi,\eta'\pi$ decay mechanisms}
The details of the $q\bar q g$ hybrid decay mechanisms represented in Figure \ref{decay_fig} are unknown.  However, the 
quark pair (one created from  the initial constituent gluon in the hybrid) can be modeled through a quark current, and
the gluon pair (one emitted from an initial constituent quark)  through a gluonic current.
Thus, for  $\pi_1\to\eta\pi,\eta'\pi$ decays, the $gg$ mechanism can be analyzed through the $\eta,\eta'$ couplings to 
the pseudoscalar gluonic current, and the $\bar q q$ mechanism through a pseudoscalar quark current.  Although it is possible to anticipate the dominance of the $gg$ mechanism in the $J/\psi$ decays and the $\bar qq$ mechanism in the $\phi$ decays because of the energy scales associated with 
the decay processes, the special role of the constituent gluon in the hybrid makes it difficult 
to make a theoretical prediction of which mechanism is dominant in hybrid decays. However, 
experimental data combined with the phenomenological analysis given below provides a means for distinguishing between these mechanisms.

If  $\pi_1\to\eta\pi,\eta'\pi$ decays  are dominated by the $gg\to\eta,\eta'$ mechanism illustrated in diagram (b) of Fig.\ \ref{decay_fig}, then we would find 
\begin{equation}
{\Gamma(\pi_1\to\eta\pi)\over \Gamma(\pi_1\to\eta'\pi)}\simeq
\left|{\langle 0|G\tilde{G}|\eta\rangle \over \langle
0|G\tilde{G}|\eta'\rangle}\right|^2\left({1-m^2_{\eta}/m^2_{\pi_1}\over
1-m^2_{\eta'}/m^2_{\pi_1}}\right)^3= 0.425\pm 0.048,
\label{gg_final_res}
\end{equation}
where $m_{\pi_1}=1.6\,{\rm GeV}$ has been used along with (\ref{GtildeG}) for the $\eta$, $\eta'$ gluonic couplings.  If the 
hybrid mass is reduced to $m_{\pi_1}=1.4$ GeV, the central value of this  ratio increases to $0.659$.

For the
 $\bar qq\to\eta,\eta'$ mechanism illustrated in diagram (a) of Fig.\ \ref{decay_fig} we find
\begin{equation}
{\Gamma(\pi_1\to\eta\pi)\over \Gamma(\pi_1\to\eta'\pi)}\simeq
\left|{\langle 0|\bar ni\gamma_5 n|\eta\rangle \over \langle
0|\bar ni\gamma_5 n|\eta'\rangle}\right|^2\left({1-m^2_{\eta}/m^2_{\pi_1}\over
1-m^2_{\eta'}/m^2_{\pi_1}}\right)^3
\end{equation}
where $m_{\pi_1}=1.6\,{\rm GeV}$ has been used, and the isovector nature of the $\pi_1$ and $\pi$ necessitates the 
non-strange quark operators in the $SU(2)$ limit.  If the theoretical value (\ref{nn_res}) is used to obtain an approximate value of the
non-strange quark operators
  we obtain 
\begin{equation}
{\Gamma(\pi_1\to\eta\pi)\over \Gamma(\pi_1\to\eta'\pi)}\approx 16 \quad,
\label{qq_final_res}
\end{equation}
which is clearly distinct from the $gg$ value (\ref{gg_final_res}).  
Even the lower bound obtained  from (\ref{ga1})
\begin{equation}
{\Gamma(\pi_1\to\eta\pi)\over \Gamma(\pi_1\to\eta'\pi)}>2.5,
\label{qq_bound_res}
\end{equation}
is sufficient to distinguish between the $gg$ and $\bar qq$ processes.
Reducing the hybrid mass to $m_{\pi_1}=1.4$ GeV increases the numerical values in 
(\ref{qq_final_res}) and (\ref{qq_bound_res})  by approximately 55\%.

In conclusion, experimental information for the decay processes $J/\psi\to\eta\gamma,\eta'\gamma$ and
$\phi\to\eta\gamma,\eta'\gamma$ has been used to 
demonstrate that these decays occur through different decay mechanisms, allowing the 
extraction  of  $\eta$, $\eta'$ couplings to gluonic and quark operators.
These extractions are consistent with those of QCD sum-rules, but have the advantage that   
they are independent  of $\eta$-$\eta'$ mixing details.   The overall consistency of these couplings substantiates the models and  
approximations used to study these decays.

Under the assumption of a hybrid nature of the $\pi_1$ states, 
the extracted couplings of $\eta$, $\eta'$ to the gluonic and quark operators are applied to estimating the decay-width ratio 
$\Gamma(\pi_1\to\eta\pi)/ \Gamma(\pi_1\to\eta'\pi)$ through the $\bar q q$ and $gg$ mechanisms illustrated in Figure \ref{decay_fig}.
This ratio is substantially different in the two mechanisms, and hence future branching-ratio measurements should identify the dominant 
decay mechanism, facilitating more detailed theoretical work. However,
we note that a sum-rule analysis in the quark scheme for $\eta$-$\eta'$ mixing clearly predicts 
$\Gamma(\pi_1\to\eta\pi)/ \Gamma(\pi_1\to\eta'\pi)>1$ \cite{tom}, suggesting  dominance of the $q\bar q$ mechanism. 
Although  effects of final-state interactions and of using a local operator 
for the $\bar q q$ and $gg$ pairs have been ignored in these processes, it seems unlikely that they will be large enough to alter the
qualitative result  $\Gamma(\pi_1\to\eta\pi)< \Gamma(\pi_1\to\eta'\pi)$ in the $gg$ mechanism and 
  $\Gamma(\pi_1\to\eta\pi)> \Gamma(\pi_1\to\eta'\pi)$ in the $\bar qq$ mechanism.
Conversely, $\pi_1$ branching ratios that lie outside the extremes
associated with (\ref{gg_final_res}) and (\ref{qq_final_res}) may be difficult to accommodate from a theoretical perspective.

\begin{acknowledgments}
Research funding from the Natural Science  \& Engineering Research Council 
of Canada (NSERC) is gratefully acknowledged. Ailin Zhang is partly 
supported by National Natural Science Foundation of China and the BEPC 
National Lab Opening Project.
\end{acknowledgments}



\begin{thebibliography}{15}
\bibitem{PDG} Particle Data Group (K. Hagiwara {\it et al} ), Phys. Rev.  {\bf D66}, 010001 (2002).

\bibitem{e852}
S. U. Chung, {\it et al.}, E852 collaboration, Phys. Rev. {\bf D60},
092001(1999);\\
G. S. Adams, {\it et al.}, E852 collaboration, Phys. Rev. Lett. 81,
5760 (1998);\\
E. I. Ivanov. {\it et al.}, E852 collaboration, Phys. Rev. Lett. 86,
3977 (2001).

\bibitem{crys}
A. Abele, {\it et al.}, Phys. Lett. {\bf B423}, 175 (1998); {\it ibid.}, 
{\bf B446}, 349 (1999).

\bibitem{bag}
R. J. Jaffe, Phys. Rev. {\bf D15}, 267 (1977);\\
M. Chanowitz and S. Sharpe, Nucl. Phys. {\bf B222}, 211 (1983);\\
T. Barnes, F. E. Close and F. de Viron, Nucl. Phys. {\bf B224},
241 (1983).

\bibitem{flux}
N. Isgur and J. Paton, Phys. Rev. {\bf D31}, 2910 (1985);\\
T. Barnes, F. E. Close,  P. R. Page and E. S, Swanson, Phys. Rev. {\bf D55}, 
4157  (1997);\\
A. Donnachie and P. R. Page, Phys. Rev. {\bf D58}, 114012 (1998);\\
P. R. Page, E. S. Swanson and A. P. Szczepaniak, Phys. Rev. {\bf D59},
034016  (1999).

\bibitem{potential} S. Godfrey and N. Isgur, Phys. Rev. {\bf D32}, 189 (1985);\\
A.P. Szczepaniak and E.S. Swanson, Phys. Rev. {\bf D55}, 3987 (1997);\\
S. Godfrey and J. Napolitano, Rev. Mod. Phys. {\bf 71}, 1411 (1999);\\
F.J. Llanes-Estrada and S.R. Cotanch, Phys. Lett. {\bf B504}, 15 (2001).

\bibitem{constituent}
D. Horn and J. Mandula, Phys. Rev. {\bf D17}, 898 (1978);\\
S. Ishida, H. Sawazaki, M. Oda and K. Yamada, Phys. Rev. {\bf D47},
179 (1993);\\
 F. Iddir, {\it et al.}, Phys. Lett. {\bf B205}, 564 (1988);\\
 F. Iddir and A. S. Safir, Phys. Lett. {\bf B507}, 183 (2001).


\bibitem{const2} F. Iddir, {\it et al.}, Phys. Lett.  {\bf B207},
325 (1988).

\bibitem{sum}
I. I. Balitsky, D. I. Dyakonov and A. V. Yung, Z. Phys. {\bf C33},
265 (1986);\\
J. Govaerts, {\it et al.}, Nucl. Phys. {\bf B284}, 674  (1987);\\
K.\ Chetyrkin and S.\ Narison, Phys.\ Lett.\ {\bf B485}, 145  (2000);\\
 H.Y. Jin, J.G. Korner, T.G. Steele, hep-ph/0211304 (to appear, Phys. Rev. D).

\bibitem{width}
F. de Viron and J. Govaerts, Phys. Rev. Lett, {\bf 53}, 2207  (1984);\\
J. I. Latorre, P. Pascual and S. Narison, Z. Phys. {\bf C34}, 347  (1987).


\bibitem{lattice}
C. Michael {\it et al.}, Nucl. Phys. {\bf B347}, 854 (1990);\\
P. Lacock {\it et al.}, Phys. Lett. {\bf B401}, 308 (1997);\\
C. Bernard, {\it et al.}, Phys. Rev. {\bf D56}, 7039 (1997);\\
C. Michael, Nucl. Phys. {\bf A655}, 12 (1999);\\
C. McNeile, {\it et al.}, Phys. Rev. {\bf D65}, 094505 (2002);

\bibitem{other}
Yu. S. Kalashnikova and Yu. B. Yufryakov, Phys. Lett. {\bf B359}, 
175 (1995);\\
P. Minkowski and W. Ochs, Phys. Lett. {\bf B485}, 139 (2000).


\bibitem{select}
F. E. Close and P. R. Page, Nucl. Phys. B443, 233 (1995);\\
P. R. Page, Phys. Lett. B401, 313(1997); ibid., B402, 183 (1997);\\
H. J. Lipkin, Phys. Lett. {\bf B219}, 99 (1989);\\
P. R. Page, Phys. Rev. {\bf D64}, 056009 (2001).



\bibitem{tom}
Ailin Zhang and T.G.\ Steele, Phys. Rev. {\bf D65}, 114013 (2002).

\bibitem{mixing}
H. Fritzsch and J. D. Jackson, Phys. Lett. {\bf 66B}, 365  (1977);\\
N. Isgur, Phys. Rev. {\bf D13}, 122  (1976);\\
F. J. Gilman and R. Kauffman, Phys. Rev. {\bf D36}, 2761  (1987);\\
H. Leutwyler, Nucl. Phys.  (Proc. Suppl.) 64, 223  (1998);\\
 T.\ Feldmann, P.\ Kroll,
     Eur.\ Phys.\ J.\ {\bf C5}, 327  (1998);\\
R. Kaiser and H. Leutwyler, hep-ph/9806336;\\
T. Feldmann, P. Kroll and B. Stech, Phys. Rev. {\bf D58}, 114006  (1998);\\
T. Feldmann, P. Kroll and B. Stech, Phys. Lett. {\bf B449}, 339  (1999).

\bibitem{pennington}
F. De Fazio and M. R. Pennington, JHEP, {\bf 0007}, 051  (2000).

\bibitem{ball}
V. A. Novikov, M. A. Shifman, A. I. Vainshtein and V. I. Zakharov, Phys.
Lett. {\bf B86}, 347  (1979);\\
P. Ball, J. M. Frere and M. Tytgat, Phys. Lett. {\bf B365}, 367  (1996);\\
E. Kou, Phys. Rev. {\bf D63}, 054027 (2001).

\bibitem{shifman}
V. A. Novikov, M. A. Shifman, A. I. Vainshtein and V. I. Zakharov, Nucl.
Phys. {\bf B165}, 55  (1980).

\bibitem{rad} P.J.\ O' Donnell, Rev.\ Mod.\ Phys.\ {\bf 53}, 673 (1981);\\
J.\ Rosner, Phys.\ Rev.\ {\bf D27}, 1101 (1983);\\
N.N. Achasov and V.N. Ivanchenko, Nucl. Phys. {\bf B315}, 465 (1989); \\
M.B.\ Cakir and G.R.\ Farrar, Phys.\ Rev.\ {\bf D50}, 3268 (1994);\\
A.\ Bramon, A.\ Grau and G.\ Pancheri, Phys.\ Lett.\ {\bf B344}, 240 (1995);\\
F.E.\ Close, G.R.\ Farrar and Zhen-ping Li, Phys.\ Rev.\ {\bf D55}, 5749
(1997);\\
E.\ Marco, S.\ Hirenzaki, E.\ Oset and H.\ Toki, Phys.\ Lett.\ {\bf B470}, 20
(1999).

\bibitem{bes}
J.Z.\ Bai {\it et al.}, BES Collaboration, Phys. Rev. {\bf D58}, 
097101 (1998).

\bibitem{kloe}
A.\ Aloisio {\it et al.}, KLOE Collaboration, Phys. Lett. {\bf B541}, 45 (2002).

\end{thebibliography}
\end{document}